\documentclass[prb,aps,twocolumn,superscriptaddress,10pt,showpacs,showkeys]{revtex4-2}
\usepackage[utf8]{inputenc}
\usepackage{amsmath}
\usepackage{bm}
\usepackage{natbib}
\usepackage{placeins}
\usepackage{graphicx}
\usepackage[separate-uncertainty=true]{siunitx}  
\usepackage[nolist]{acronym}
\usepackage{color}
 \usepackage{hyperref}
\hypersetup{colorlinks=true, citecolor=blue, urlcolor=blue, linkcolor=blue}

\begin{document}
\title{Ultrafast electron dynamics upon above band-gap excitation in epitaxial LaFeO$_3$(001) thin films}

\author{Friederike Elisa Wührl}
\email{friederike.wuehrl@physik.uni-halle.de}
\author{Antonia Rieche} 
\author{Anne Oelschläger}
\author{Kathrin Dörr}
\author{Wolf Widdra}
\affiliation{Institute of Physics, Martin-Luther-Universit\"at
Halle-Wittenberg, D-06120 Halle, Germany}

\date{\today}

\begin{abstract} 
Strong electron correlations in perovskite oxides give rise to rich and often unexpected electronic phenomena. In this study, we present a comprehensive surface-science investigation of epitaxial thin films of the charge-transfer insulator LaFeO$_3$(001). The characterization includes low-energy electron diffraction (LEED), high-resolution electron energy loss spectroscopy (HREELS), and photoemission spectroscopy. We map both the occupied and unoccupied electronic states using two-photon photoemission (2PPE) spectroscopy. Furthermore, we probe electron dynamics through an ultraviolet-ultraviolet (UV-UV) pump-probe experiment, exciting electrons from hybridized O~$2p$/Fe $3d$ states to Fe minority-spin states above the band gap. Our results reveal three distinct unoccupied states, which we assign to Fe $t_{2g\downarrow}$, Fe $e_{g\downarrow}$, and La $5d$ orbitals. Notably, the conduction band minimum exhibits a biexponential decay with time constants of 39\,fs and 1100\,fs, suggesting the presence of two independent decay pathways.
\end{abstract}

\keywords{charge-transfer insulator, LaFeO$_3$, time-resolved photoexcitation}

\maketitle   

\section{Introduction}

Perovskite oxides, with the general chemical formula ABO$_3$, represent a remarkably versatile class of materials that exhibit a wide range of emergent electronic and magnetic phenomena, including high-temperature superconductivity, colossal magnetoresistance, multiferroicity, and metal-insulator transitions~\cite{imada1998, tokura2000, dagotto2005, hwang2012}. This diversity arises from the ability to finely tune their structural, electronic, and magnetic properties through chemical substitution, epitaxial strain, dimensional confinement, and defect engineering~\cite{mannhart2010, zubko2011}. As a result, perovskite oxides are of great interest both fundamentally, for exploring strongly correlated electron systems, and technologically, for applications in oxide electronics, catalysis, and spintronics.

A central factor determining the ground state and excitations of these materials is the degree of electron-electron correlation, particularly in the narrow $3d$ bands of transition metal ions. When the on-site Coulomb repulsion $U$ exceeds the electronic bandwidth $W$, a metal-to-insulator transition occurs, and the system becomes a Mott insulator. Within the Zaanen-Sawatzky-Allen (ZSA) classification scheme~\cite{zaanen1985}, transition metal oxides can be further divided into two distinct classes based on the relative magnitudes of $U$ and the charge-transfer energy $\Delta$: Mott-Hubbard insulators ($\Delta > U$), where the band gap opens between the lower and upper Hubbard bands of the $3d$ electrons, and charge-transfer insulators ($\Delta < U$), where the gap lies between the oxygen $2p$ ligand states and the unoccupied upper Hubbard band.

\begin{figure}
\includegraphics[width=\columnwidth]{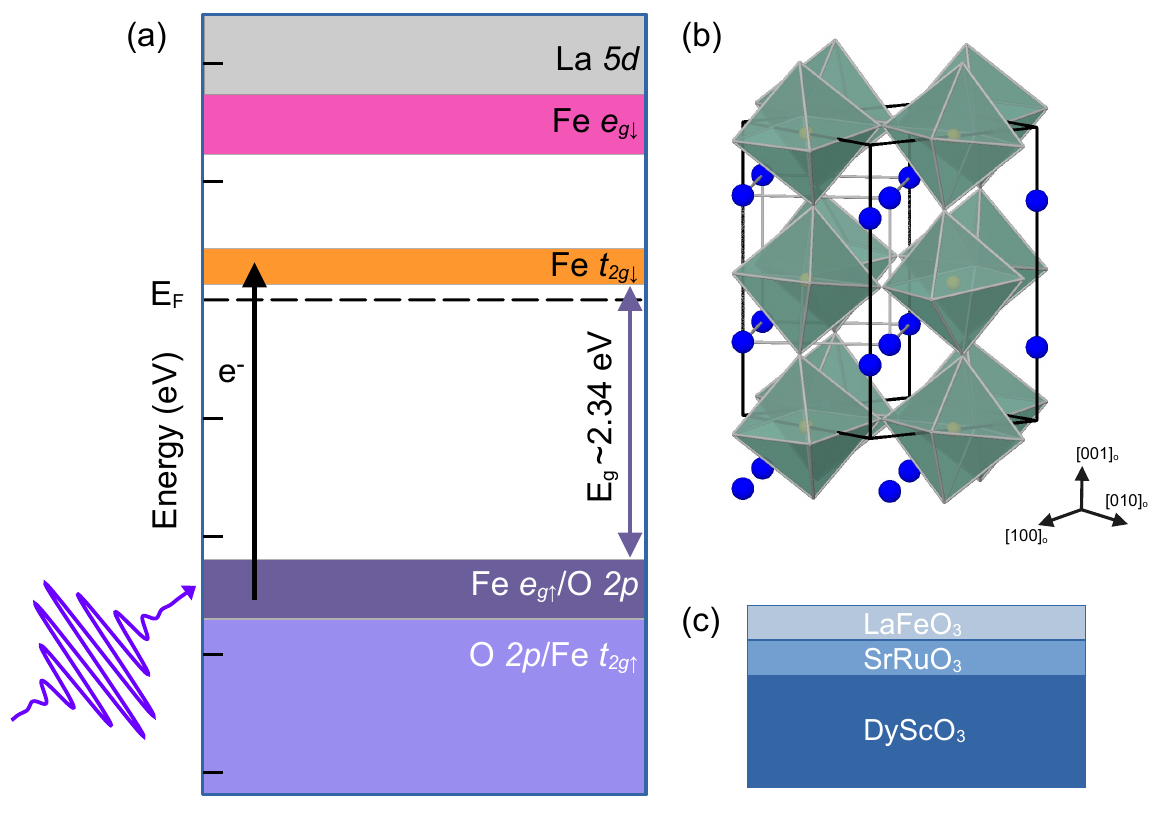}
\caption{(a) Schematics of bandstructure according to Scafetta \textit{et al.} \cite{scafetta2014}. The Fermi level has been aligned according to experimental photoemission data. The notation Fe $e_{g\uparrow}$ refers to a single atom. The opposite spin channel is found for LaFeO$_3(001)$  as AFM at the neighboring Fe atom  \cite{scafetta2014}. (b) Orthorhombic unit cell with a$^{-}$a$^{-}$c$^{+}$ rotation pattern. (c) Sample layout: three layer system: thin film of LaFeO$_3(001)$ on orthorhombic DyScO$_3(110)_o$ with a metallic SrRuO$_3$ film in between. SrRuO$_3$ acts as back electrode for photoemission experiments.}
\label{Abb_0_Introduction}
\end{figure}

LaFeO$_3$ is a prototypical example of a charge-transfer insulator. It crystallizes in a slightly orthorhombically distorted perovskite structure (space group \textit{Pnma}) (Fig.\,\ref{Abb_0_Introduction}(b)) with an a$^{-}$a$^{-}$c$^{+}$ rotation pattern in Glazers notation \cite{Glazer1972, zhu2024}  and features a nominal Fe$^{3+}$ ($3d^5$, high-spin) configuration. The Fe $3d$ states exhibit strong hybridization with the O $2p$ orbitals, leading to a band gap of approximately \SI{2.3}{eV}~\cite{Arima1993}. Beyond its electronic structure, LaFeO$_3$ is known for its robust G-type antiferromagnetism, in which each Fe$^{3+}$ spin is antiparallel to all six nearest neighbors. The material exhibits the highest N\'eel temperature among the orthoferrites of approximately \SI{740}{K}~\cite{Selbach12}, indicating strong superexchange interactions mediated through the Fe--O--Fe network. This magnetic ordering plays a significant role in influencing the electronic band structure and spin-dependent dynamics of excited carriers.

Despite its insulating nature, LaFeO$_3$ supports rich photoexcitation dynamics that are only beginning to be fully understood. Time-resolved photoemission spectroscopy, particularly in the form of two-photon photoemission (2PPE) and pump-probe techniques, has emerged as a powerful method to investigate both the occupied and unoccupied states, as well as the ultrafast relaxation processes of photoexcited electrons and holes~\cite{petek1997, rohwer2011, hellmann2012}. In such experiments, an initial pump pulse excites electrons across the band gap, and a subsequent probe pulse maps the transient population of intermediate states with femtosecond temporal resolution. This allows direct access to excited-state lifetimes, carrier relaxation pathways, and electron-phonon or electron-electron scattering dynamics --- information that is crucial for developing microscopic models of correlated oxides and for designing materials for optoelectronic applications. To date, time-resolved two-photon photoemission (2PPE) studies on perovskites have been primarily focused on metal halides \cite{rieger2023, lin2022a}, attributed to their exceptionally long lifetimes and relevance in the solar cell industry; no studies have been reported for perovskite oxides.

The simplified schematics of the LaFeO$_3$(001) bandstructure is given in Fig.\,\ref{Abb_0_Introduction}(a), which is based on DFT calculations of Scafetta \textit{et al.} in Ref. \onlinecite{scafetta2014}. Due to the low band dispersion, we have omitted the k-dependence in the graph. The conduction band is characterized by Fe minority states and a secondary band gap is expected between the  $t_{2g\downarrow}$ and $e_{g\downarrow}$ states. The valence band is dominated by oxygen $2p$ levels with a strong hybridization to Fe majority $3d$ states. As LaFeO$_3$ is a G-type antiferromagnet, the opposite spin channel is found at neighboring Fe ion. 
Due to the lack of inversion symmetry, the Dzyaloshinskii--Moriya interaction induces a slight canting of the antiferromagnetic spins, resulting in a weak ferromagnetic moment~\cite{moriya1960}. 

The unoccupied states have been studied so far only with X-ray absorption spectroscopy, optical absorption spectroscopy \cite{scafetta2014, Arima1993}, and inverse photoemission spectroscopy \cite{chainani1993}, mainly focusing on changes after substitution of La to Sr. 

In this work, we employ a combination of surface-sensitive techniques---including low-energy electron diffraction (LEED), high-resolution electron energy loss spectroscopy (HREELS), and angle-resolved photoemission spectroscopy (ARPES)---to investigate the electronic structure of epitaxial LaFeO$_3$(001) thin films. Using two-photon photoemission (2PPE) and UV-UV pump-probe spectroscopy, we further map the unoccupied states and explore the ultrafast dynamics of photoexcited electrons. Our findings offer new insights into the nature and relaxation pathways of electronic excitations in a canonical charge-transfer insulator.

\section{Methods}
Thin films of LaFeO$_3$(001) have been grown by pulsed laser deposition (PLD) using a KrF excimer laser with a wavelength of $248$\,nm on DyScO$_3$(110)$_o$ substrates (\textit{Crystec GmbH}). The lattice mismatch between pseudocubic LaFeO$_3(001)$ and DyScO$_3$(110)$_o$ amounts to $0.4$\%, which enables epitaxial growth with high quality \cite{dixon2015}. 
 Firstly, a SrRuO$_3$ layer was grown on DyScO$_3$(110)$_o$ substrates, followed by the LaFeO$_3$(001) layer. Thereby 25\,nm SrRuO$_3$ acts as back electrode for photoemission experiments, bulk samples would lead to strong charging effects in photoemission experiments.  The preparation parameters have been 970\,K, an oxygen pressure of \SI{0.1}{mbar} for LaFeO$_3$ and $0.2$\,mbar for SrRuO$_3$ and an energy density of $1$\,J/cm$^2$. After growth, the samples are cooled down to room temperature under an oxygen pressure of $200$\,mbar, subsequently transferred via air to an ultrahigh vacuum system with a base pressure of $10^{-10}$\,mbar and  cleaned from adsorbates through UHV heating. They have been characterized with low energy electron diffraction (LEED), high resolution  electron energy loss spectroscopy (HREELS, Delta $0.5$, SPECS GmbH), ultraviolet photoemission spectroscopy (UPS), X-ray photoemission spectroscopy (XPS) and two-photon photoemission (2PPE). The XPS-spectra of Fe $2p$, La $3d$ and O $1s$ are in accordance with literature \cite{Oh1988, wadati2005, wang2019a} and confirm the Fe$^{3+}$ and La$^{3+}$ oxidation states. Additionally, air transfer leads to a small carbon contamination. Exposure of the sample to X-rays leads to band bending, evident from a shift of the 2PPE peak by 150\,meV towards the Fermi level as well as small 1PPE signal due to defect states below $E_F$.  This process was fully reversible by heating the sample to 770\,K in UHV, which probably enables oxygen diffusion from the bulk to the surface.  The energy $E$ and angle $\theta$ of the emitted photoelectrons  are analyzed simultaneously using a Phoibos 150  hemisphere (SPECS GmbH) with a 2D CCD detector. The limitation to measure along one high symmetry direction is overcome by the possibility of azimuthal rotation of the sample. 
 The photoelectrons are excited through a standard discharge lamp using the He I line for UPS, and a tunable fs laser system, allowing for 2PPE with photon energies of 1.8 to 3.4\,eV  and 3.4 to 4.7\,eV for pump and probe. The laser setup consists of a double nonlinear optical parametric amplifier which is pumped by a 13\,W fiber laser (IMPULSE, Clark-MXR) at $1.2$\,MHz repetition rate. Both outputs of the NOPA system can be frequency doubled, allowing for both UV-IR and UV-UV pump-probe measurements \cite{hofer2011a, gillmeister2020}. Typical laser pulse energies have been between $1-$\SI{20}{\mu J\per\square\cm}.
\section{Results}

\subsection{LEED}

\begin{figure}
\includegraphics[width=\columnwidth]{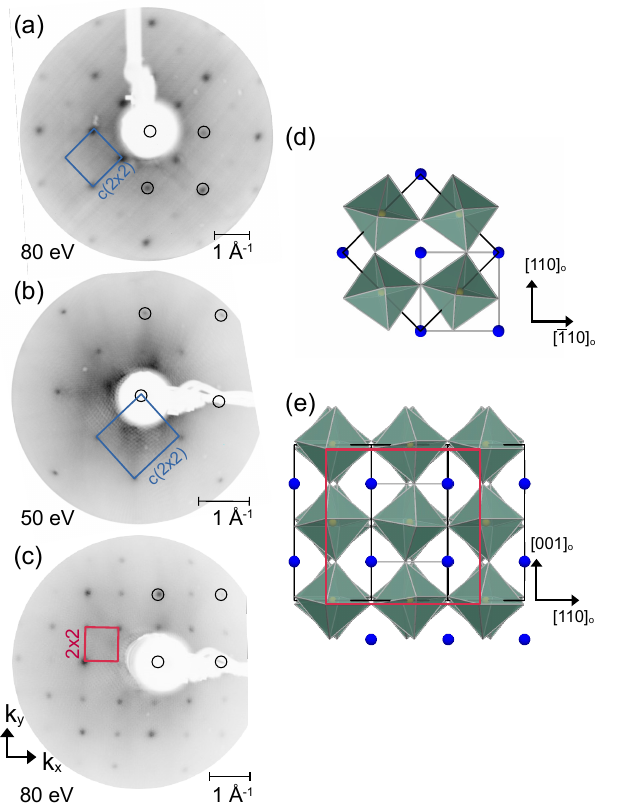}
\caption{(a)-(c) LEED images for three different LaFeO$_3$(001) thin film samples. The spots according to the pseudocubic unit cell are marked with black circles, the c(2x2) superstrucutre blue and the (2x2) superstructure red. 
(d) LaFeO$_3$(001)$_o$ surface: the rotation of the oxygen octahedra around the c-axis leads to a c(2x2) superstructure according to the pseudocubic unit cell. (e) LaFeO$_3$(110)$_o$ plane: the in-phase rotation of the oxygen octahedra around the c-axis and the out of phase rotation around the two other pseudocubic axis lead to a (2x2)-superstructure.}
\label{Abb_1_LEED}
\end{figure}
The low energy electron diffraction (LEED) images of three PLD-grown films are shown in Fig.\,\ref{Abb_1_LEED}(a)-(c). The sample in Fig.\,\ref{Abb_1_LEED}(a) has been grown with a LaFeO$_3$ film thickness of 15\,nm, whereas the other samples in (b) and (c) have thicknesses of  7\,nm.  Sample (b) has been transferred through a vacuum suitcase system, avoiding exposure to air, whereas samples (a) and (c) have been transferred via air. All films have been cleaned from adsorbates by final annealing in UHV to  $700-900$\,K. The sharp diffraction spots in all three LEED images confirm the formation of epitaxial single-crystalline layers.  The (1x1) diffraction spots, which correspond to the unreconstructed surface structure of the pseudocubic unit cell of the perovskite structure, are marked with black circles. However, we observe additional and different superstructures for sample (c) in comparison with samples (a) and (b). The superstructure diffraction spots can be either described as c(2x2) structure (blue) for samples (a) and (b) or as (2x2) superstructure (red) for (c). We attribute the superstructures to the rotation of oxygen octahedra in the orthorhombic perovskite structure, as this alters the symmetry from cubic to orthorhombic \cite{moore2007,siwakoti2021,kyung2021,hu2010,park2020}. Such a symmetry reduction is present for all orthoferrites, whereas LaFeO$_3$ shows the lowest deviation from a cubic structure with the smallest rotation angles for the oxygen octahedra \cite{weber2016}. Two different orthorhombic orientations with respect to the LaFeO$_3$(001) surface are sketched in Fig.\,\ref{Abb_1_LEED}(d) and (e). In Fig.\,\ref{Abb_1_LEED}(d) the (001)$_o$ plane is shown, where the index o emphasizes the orthorhombic nomenclature and where the orthorhombic c-axis is oriented along  the surface normal. The alternating octahedra rotation around the c-axis results in a c(2x2) superstructure with respect to the pseudocubic unit cell. Additionally, glide mirror planes in [010]$_o$ and [100]$_o$ directions lead to extinction of every second diffraction spot along these two directions \cite{siwakoti2021}. 
An additional rotation around the two other pseudocubic axis, from the (001)$_o$ surface one would consider as tilt, reduces the number of glide planes to one. This has been observed for various systems with octahedral rotations in literature \cite{moore2007, hu2010}. In our case, the LEED images of the LaFeO$_3$ film in (a) and (b) show the c(2x2) superstructure, but missing out any extinctions due to glide mirror planes. This can be explained by domains of different rational directions \cite{siwakoti2021, li2013}. We conclude, that the absence of glide mirror planes confirms the presence of rotation angles around all three pseudocubic axis. 
The second case is shown in Fig.\,\ref{Abb_1_LEED}(e) for the (110)$_o$ plane, where the orthorhombic c-axis is 90° rotated with respect to the surface normal. Here, the in-phase rotation of the octahedra around the c-axis and the out of-phase rotation around the two other pseudocubic axis results in a (2x2) superstructure, which is in accordance with the observed LEED image of LaFeO$_3$ in (c).   
 Additionally, we validate this expectation by the observation of a (2x2) LEED superstructure for the bare surface of the orthorhombic DyScO$_3$(110)$_o$ substrate (\textit{Crystec GmbH}), which is known to grow with orthorhombic c-axis orientation in-plane (90° rotated with respect to the substrate surface). 
For the interpretation of Fig.\,\ref{Abb_1_LEED}(c), however, a superposition with a c(2x2) superstructure can not be excluded. Further investigations, as e.g. by a LEED I(V) characterization, are required to separate both. In summary, we have observed LaFeO$_3$(001)  films with orthorhombic c-axis oriented along the surface normal for thick films and for ultrathin films transferred under UHV conditions, Fig.\,\ref{Abb_1_LEED}(a) and (b). Whereas the surface structure of ultrathin films transferred through air show a mixed orthorhombic domain structure, Figs.\,\ref{Abb_1_LEED}(c). This might also hint to slight stoichiometric variations between the samples as a Fe deficit already leads to a reduced c-axis parameter \cite{scafetta2017}.

The following characterization, concerning the phonon (HREELS) and electronic properties (2PPE, UPS) has been done on the third sample (Fig.\,\ref{Abb_1_LEED}(c)). Nevertheless we do not expect that the surface plane affect the electronic properties. 

 \subsection{HREELS}
\begin{figure}
	\includegraphics[width=0.8\columnwidth]{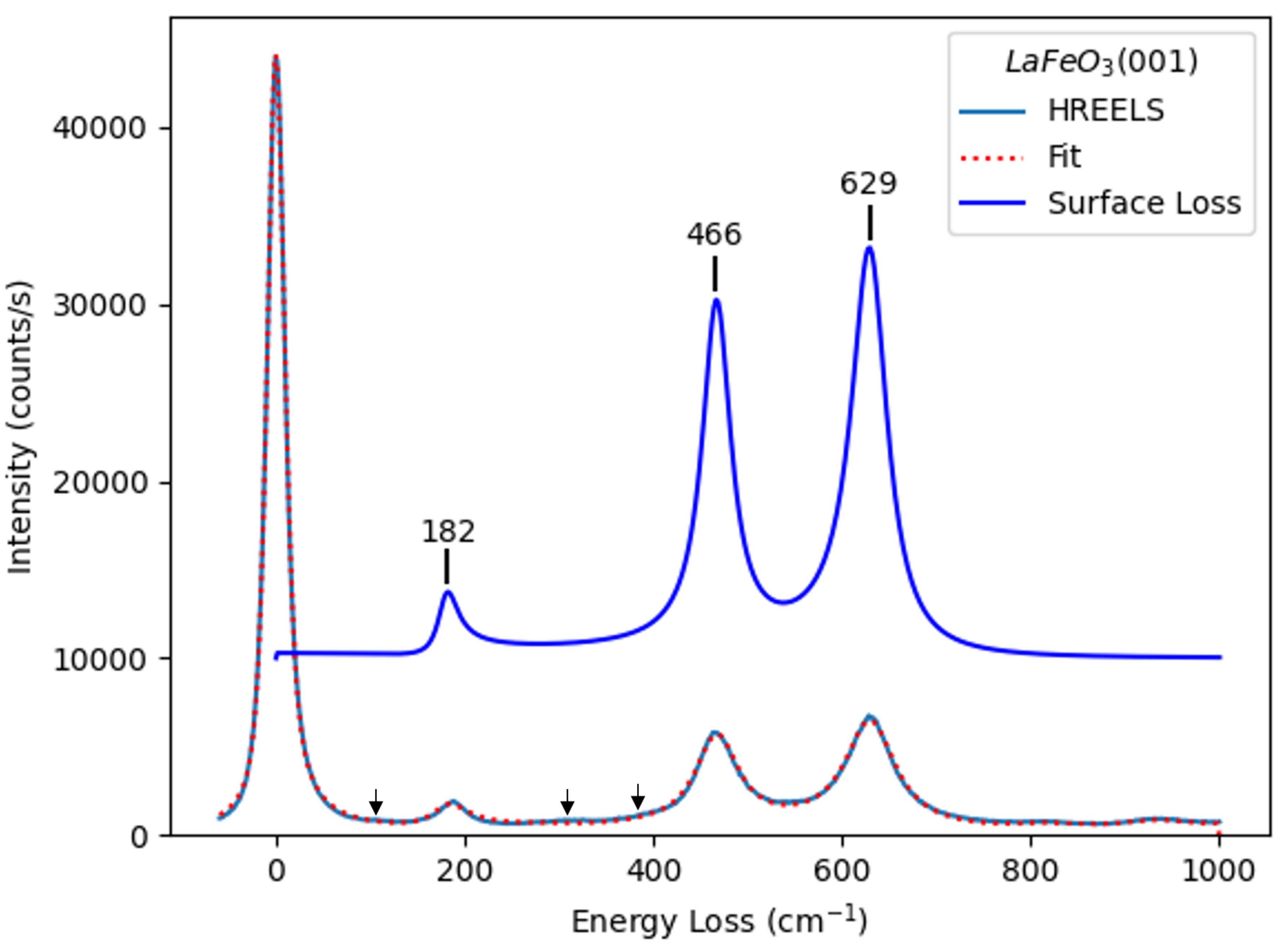}
	\caption{HREELS spectrum of LaFeO$_3$(001) thin film (bottom line) for specular electron scattering with primary energy of 4\,eV and angle of incidence of 55°. Fitted energy loss spectrum (red dashed line) based on three surface phonon polaritons. Extracted surface loss spectrum (blue line), vertically offset by 10000 counts/s. Small vertical black arrows indicate energetic positions of phonon contributions from the cubic-to-orthorhombic symmetry reduction.}
	\label{Abb_HREELS}
	\end{figure}

 The LaFeO$_3$(001) thin film has been characterized by high-resolution electron energy loss spectroscopy (HREELS) as depicted in Fig.\,\ref{Abb_HREELS} to address the characteristic surface phonon properties of a perovskite oxide. The HREELS measurements have been performed in a separate UHV system with a base vacuum of $10^{-10}$\,mbar. A primary electron energy of \SI{4}{eV} is used with a total energy resolution of $2.0-2.5$\,meV ($16-$\SI{20}{\per\cm}), depending on the long-range surface order \cite{premper2020,kostov2011, kostov2016}. 
The surface of the LaFeO$_3$(001) thin film is characterized by three prominent perovskite surface phonon polaritons, which are characteristic for the oxide perovskite as has been shown earlier for BaTiO$_3$(001) thin films \cite{premper2020}. The HREELS spectrum of LaFeO$_3$(001) in Fig.\,\ref{Abb_HREELS} (bottom curve) has three surface phonon polaritons at 629, 466, and \SI{182}{\per\cm} at $k_\parallel=0$ and can be quantitatively fitted as indicated by the dashed line. These polariton modes correspond to longitudinal optical phonons at 670, 479, and \SI{182}{\per\cm} in the LaFeO$_3$(001) thin film.
In the derived surface loss spectrum (blue solid line) the finite instrument function and multiple scattering effects are removed and the narrow phonon-polariton linewidths (full width of half maximum) can be determined to 44, 38 and \SI{31}{\per\cm} for the surface phonon polaritons at 629, 466, and \SI{182}{\per\cm}. The dominance of three dipole-active phonon-polaritons stems from a nearly cubic structure. Small orthorhombic deviations lead to new weak modes at 110, 304, and \SI{385}{\per\cm} at locations indicated by vertical arrows but are barely visible. Note that the absence of additional loss peaks excludes the presence of other competing oxide structures.

 \subsection{Electronic structure}
\begin{figure*}
\includegraphics[width=\textwidth]{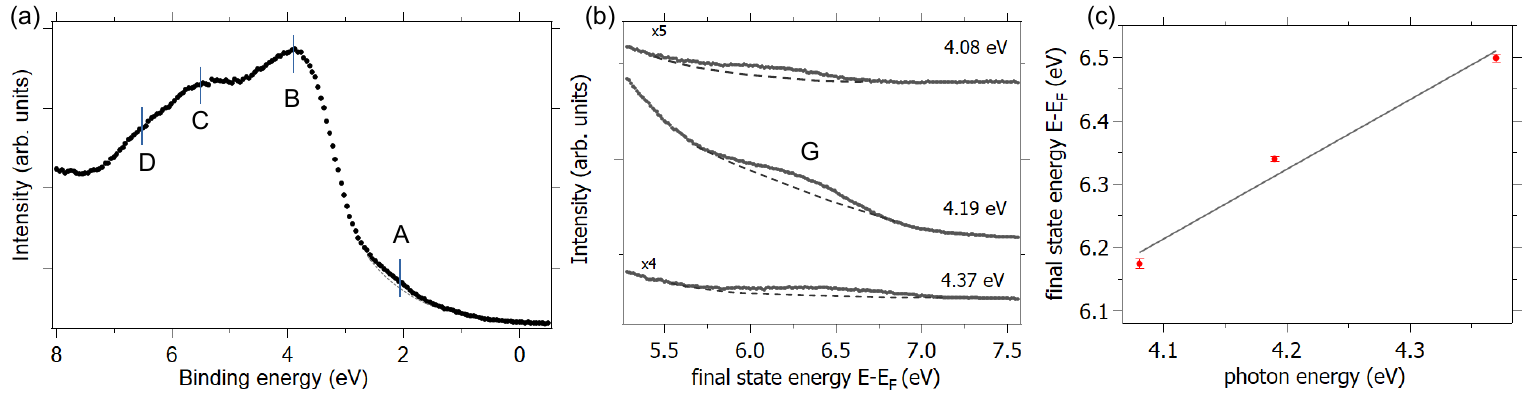}
\caption{Photoemission experiments: (a) UPS for LaFeO$_3(001)$ ($h\nu=21.2$\,eV, k$_{\text{x}}$ direction). (b) 2PPE on LaFeO$_3$ with varying photon energies at time delay zero between pump and probe pulse. The labeled photon energies apply to both, pump and probe energies, vertical offset against each other (k$_{\text{x}}$ direction at \SI{0.17}{\per\angstrom}). (c) Photon energy dependence of final state energy $E-E_F$ extracted from (b). The slope is determined to be $1.1\pm0.1$.}
\label{Abb2_UPS_2PPE}
\end{figure*}

The UPS spectrum, probing the valence band structure of LaFeO$_3$ is shown in Fig.\,\ref{Abb2_UPS_2PPE}(a). It was measured at the $\overline{\Gamma}$ point, with an acceptance angle of $\pm 7^{\circ}$. The work function was determined to be 4.4\,eV from the secondary electron cut off. We find the onset of photoemission at 0.4\,eV with a strong rising edge towards the two main peaks at 3.9 and 5.4\,eV, which are labeled with B and C in the graph. Additionally we assign the shoulder at 6.4\,eV to state D. On the rising edge one finds another feature A with low spectral weight. The dashed line serves thereby as guide for the eye for the rising edge. Angle resolved data did not give any indications for dispersive features over $k_{\parallel}$. This spectra can be compared with Wang   \textit{et al.} \cite{wang2019a} and Wadati \textit{et al.} \cite{wadati2006}. They find three states A, B, D at 1.9\,eV Wang (2.1\,eV Wadati), 3.4\,eV (3.9\,eV) and 6.1\,eV (6.7\,eV) probing the valence band structure in the X-ray regime at 800\,eV and 1486\,eV. The slight deviations in the splitting between the states  and the varying spectral weights, suppressing the peak C strongly in XPS-valence band photoemission, are probably a consequence of different photon energies used, which probe the Brillouin zone at distinct points.  Wadati \textit{et al.} \cite{wadati2006} performed element resolved tight binding calculations for the density of states and showed that B and D are O $2p$ derived states with strong hybridization to the Fe $3d$ states, whereas  A is assigned to Fe $e_{g\uparrow}$ states.  \par
2PPE spectra for photoexcitation across the band gap are displayed in Fig.\,\ref{Abb2_UPS_2PPE}(b). The spectra were measured at zero time delay between pump and probe pulse, whereas both pulses have the same photon energy. The spectra have been recorded for $k_x$ = \SI{0.17}{\per\angstrom} with an angle of light incidence of $50^{\circ}$ with respect to the surface normal. The spectra have been averaged over $\pm 10^{\circ}$ .  2PPE spectra for three different photon energies are plotted with an offset against each other. The dashed lines are guide for the eyes for the secondary electron background.  At all three photon energies we find a state, labeled G, which shifts with the photon energy.  For $h\nu=4.19$\,eV its position is at a final state energy of $6.31$\,eV with a FWHM of $0.55$\,eV. 
In Fig.\,\ref{Abb2_UPS_2PPE}(c), the position of the final state energy of state G with respect to the photon energy is plotted; the determined slope of  ($1.1\pm0.1$) points out that the final state energy depends only on the probe pulse, the state F is therefore assigned to a state at 2.1 eV above E$_F$. 

\begin{figure*}[t]
\includegraphics[width=\textwidth]{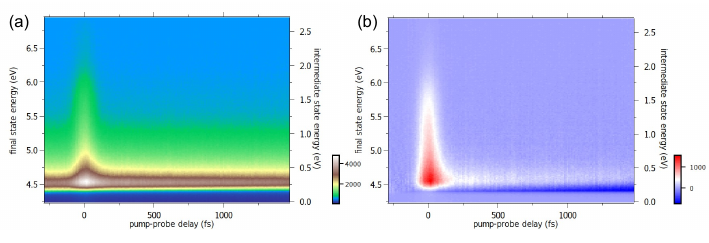}
\caption{Time-resolved 2PPE on LaFeO$_3$(001) with photon energies of $h\nu_{\text{pump}}=3.43$\,eV, $h\nu_{\text{probe}}=4.24$\,eV, p-p polarized. (a) Unprocessed data. The energy scale is given as final and intermediate state energy. Color Scale given in counts. (b) Background subtracted 2PPE spectra. The color scale encodes the electronic population caused by the two-color pump and probe pulse.}
\label{Abb4_tr2PPE}
\end{figure*}
\begin{figure*}
\includegraphics[width=\textwidth]{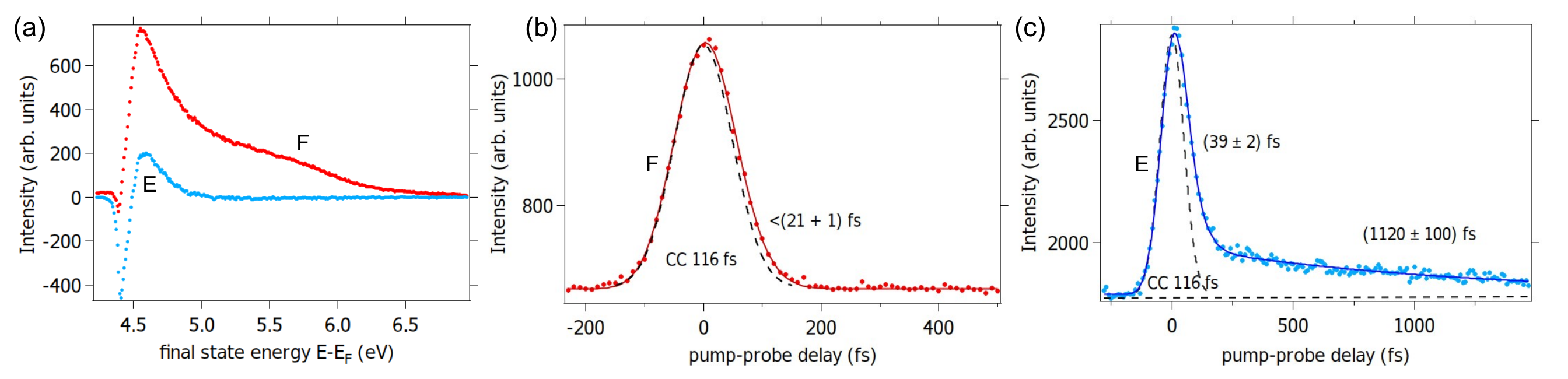}
\caption{(a) Energy distribution curves extracted from Fig.\,\ref{Abb4_tr2PPE}(b) for short delay times (red line, integrated from -150 to 180\,fs) and for longer delay times (blue line, integrated from 200 to 1500\,fs).  (b) Decay dynamics of state F, integrated over the final state energies of 5.2 to 6.2\,eV. (c) Decay dynamics of state E, integrated over the final state energies from 4.6 to 4.9\,eV. The solid lines represent the fitted data for a convolution of the pump-probe cross correlation and a single (b) or biexponential decay (c). The resulting lifetimes are given in the graph. The pump-probe cross correlation is displayed as dashed line in both graphs.  }
\label{Abb5_2PPE_EDCandLDC}
\end{figure*}
\par 
To explore the lifetimes of the unoccupied states the time delay of the pump  ($h\nu_{\text{pump}}=3.43$\,eV)  and probe pulse ($h\nu_{\text{probe}}=4.24$\,eV) is varied. The resulting color-coded energy-resolved two dimensional data in dependence of the pump-probe delay are shown in Fig.\,\ref{Abb4_tr2PPE}(a).  At time zero, an additional signal is observed due to the pump-probe overlap, however the 2D plot is dominated by monochromatic two-photon photoemission (2PPE) intensity at positive and negative time delays, a result of the UV range of both pulses. The background subtracted 2PPE spectrum is shown in Fig.\,\ref{Abb4_tr2PPE}(b), where the  monochromatic 2PPE intensity was subtracted using the energy distribution curve at negative time delays, at which the probe pulse arrives before the pump pulse. The energy distribution curves (EDC), extracted from Fig.\,\ref{Abb5_2PPE_EDCandLDC}(b) and integrated for time delays between -150 and 180\,fs (red) and between 200 and 1500\,fs (blue) are compared in Fig.\,\ref{Abb5_2PPE_EDCandLDC}(a). The negative signal observed at 4.4\,eV is an artefact resulting from the background subtraction, as the work function of the sample changed slightly by 100\,meV during the time-resolved experiment, causing a shift of the secondary electron cut off to higher energy values. This shift is also evident in the background-subtracted 2D plot as a blue line at 4.4\,eV final state energy. The red EDC shows two contributions, a peak at $E-E_F=5.68$\,eV (state F) with a FWHM of 0.64\,eV and a second component (E) which appears at lower energies directly above the secondary electron cut off and is superposed with the secondary electron background. The second component gets more evident in the blue EDC in comparison, which integrates over the contribution for long  pump-probe delays, whereas state F shows no intensity at longer pump-probe delays. For these two states we have extracted the lifetimes, taking line profiles of time-resolved 2PPE spectra across their energy range between $5.2$ and $6.2$\,eV and between 4.6 and 4.9\,eV, respectively,  as is depicted in Figs.\,\ref{Abb5_2PPE_EDCandLDC}(b) and (c). Lifetimes have been extracted using a rate-equation model fitting the data as a convolution of cross correlation with an exponential decaying function. The cross correlation (CC) with a width of $116$\,fs, which has been determined from a virtual state at final state energies of $6.3$ to $6.5$\,eV, sets a lower limit of  $18$\,fs for the lifetime determination. State F shows a short lifetime of $<21+1$\,fs, already near the resolution limit of our experiment, whereas state E reveals a biexponential decay with two time constants ($39\pm 2$)\,fs and $(1.1\pm0.1)$\,ps. 
\section{Discussion}
In this section, we will discuss the assignment of the observed peaks in 2PPE to unoccupied states and their alignment in the energy level diagram, referring to Fig.\,\ref{Abb_0_Introduction}(a) again.
The Fermi level is located 0.3\,eV below the conduction band minimum. This positioning indicates that the system is n-doped, which is consistent with data in other publications considering the peak position of the O $2p$-derived valence band states B, C, D in valence band photoemission \cite{wadati2006,wang2019a}. The unoccupied states E, F, and G that have been identified in 2PPE will be assigned in the following to the specific electronic states Fe~$t_{2g\downarrow}$,  Fe~$e_{g\downarrow}$, and La~$5d$, respectively \cite{scafetta2014, wadati2005}. 

The two-color 2PPE experiments (Fig.\,\ref{Abb4_tr2PPE} and Fig.\,\ref{Abb5_2PPE_EDCandLDC}) identified the broad unoccupied state F at 1.4\,eV above $E_F$ with a lifetime of $<21$\,fs. It is assigned to the Fe $e_{g\downarrow}$ state based on the calculation of Ref.\,\onlinecite{scafetta2014}. In these experiments, the state F has been photoexcited from the occupied state A (Fe $e_{g\uparrow}$) via a pump photon energy of 3.43\,eV. This spin forbidden transition is  likely permitted due to d-p hybridization  \cite{scafetta2014}. In the one-color experiments (Fig.\,\ref{Abb2_UPS_2PPE}(b)) with photon energies between 4.08 and \SI{4.37}{eV} it is not present. The required initial state (at about 2.9\,eV below $E_F$) would correspond to the rising edge of B and C. From symmetry arguments this transition would be allowed. However, it is not observed in accordance with optical absorption measurements and calculations, which also did not identify a transition originating from this region of the valence band \cite{scafetta2014}. 
The unoccupied state E located at \SI{0.3}{eV} just above $E_F$ is assigned to Fe $t_{2g\downarrow}$. In the two-color 2PPE experiments (Fig.\,\ref{Abb4_tr2PPE}) with a pump energy of \SI{3.43}{eV}, it is photoexcited from an initial state at about $-3.1$\,eV. We conclude that laser excitation from the 2.9 to \SI{3.1}{eV} region below $E_F$ is only possible into state E but not into state F. Indirect population of state E could be however imaginable. To prove this, further systematic variations in pump and probe energies will be necessary. The observed splitting of the unoccupied Fe $t_{2g\downarrow}$ and Fe $e_{g\downarrow}$ states of $\sim1$\,eV is slightly smaller than reported in XAS measurements. It might be a consequence of different final states, since in XAS one creates a hole in the oxygen K shell.

In the one-color measurements of Fig.\,\ref{Abb2_UPS_2PPE}(b) and (c), we have identified the  unoccupied state G at 2.1\,eV above the Fermi energy. There, a pump energy of \SI{4.19}{eV} photoexcited the electron from the initial state A (Fe $e_{g\uparrow}$), which is located about \SI{2.1}{eV} below $E_F$. However, in the time-resolved 2PPE data (Fig.\,\ref{Abb4_tr2PPE}) with different pump-probe energies we do not observe it as separate peak. With photon energies of 3.43 and $4.24$\,eV, a photoexcitation across the band gap into state G is only possible with a pump energy of $4.24$\,eV, which leads to a probe final state energy of 5.5\,eV. This final state energy is just 100\,meV below the final state energy of state F. However, a finite lifetime of state G would extend to negative delay times in Figs.\,\ref{Abb4_tr2PPE} and Fig.\,\ref{Abb5_2PPE_EDCandLDC} since the role of pump and probe photons are interchanged here. Therefore we assume that the final state in the region of 5.2 to 6.2\,eV is a superposition of photoemission from states F and G, whereas the state G (La $5d$) has a lifetime below the resolution limit of our experiment.  

 The lifetimes of electron population in states F and G (Fe $e_{g\downarrow}$ and La $5d$) are exceptionally short. The short lifetime of La $5d$ may rise from fast electron-electron scattering and thermalization because of its large bandwidth and high energetic position \cite{haight1995}. For state F we would have expected a longer lifetime because of the small band gap between Fe $t_{2g\downarrow}$ and Fe $e_{g\downarrow}$. There seems to be an efficient decay channel, on a similar time scale as it was observed for NiO \cite{gillmeister2020}. However, a specific relaxation pathway could not be unraveled in our study. For the conduction band minimum, state E (Fe $t_{2g\downarrow}$), two time constants are observed with nearly an order of magnitude difference. Two independent decay channels have to be present here. There are several possible explanations: Typical for semiconductors is electron-hole recombination on timescales of picoseconds, which can be accelerated by defect assisted recombination. We have no indications for defects in the band gap as no 1PPE signal was observed. Another proposal is given by Nitzchke \textit{et al.} \cite{nitschke2025} for the Mott insulator FePS$_3$. They attribute the observed biexponential decay with the two time constants to spin-forbidden and spin-allowed pathways, whereas the spin-allowed is mediated through the neighboring Fe ion with opposite spin.

\section{Summary}
We have grown epitaxial LaFeO$_3$(001) thin films with high surface quality on SrRuO$_3$/DyScO$_3$(110). The observed (2$\times$2) and c(2$\times$2) LEED superstructures, relative to the pseudocubic unit cell, indicate symmetry lowering due to octahedral rotations. HREELS measurements revealed three characteristic surface phonon polaritons, consistent with a near-cubic perovskite oxide structure. Time-resolved two-photon photoemission spectroscopy identified three unoccupied electronic states located 0.3, 1.4, and 2.1\,eV above the Fermi level, which we attribute to Fe $t_{2g\downarrow}$, Fe $e_{g\downarrow}$, and La $5d$ states, respectively. A band gap of 2.3\,eV was determined, with the Fermi level situated 0.3\,eV below the conduction band minimum, indicating an n-type insulating character. The lifetimes of these unoccupied states varied significantly, with the conduction band minimum exhibiting two distinct decay components of 39\,fs and 1.1\,ps, while the higher-energy states decayed within 21\,fs and below 18\,fs. While some questions remain open, our results uncover intriguing aspects of electron dynamics in LaFeO$_3$ and establish a foundation for exploring strongly correlated perovskite oxides. This approach may be extended to related systems, such as BiFeO$_3$, where the substitution of La with Bi introduces a ferroelectric component.

\section*{acknowledgement}
We thank R. Kulla for technical support. Financial support by the Deutsche Forschungsgemeinschaft (DFG) through the collaborative research center SFB 227 (Ultrafast Spin Dynamics, projects A05 and A06).

\end{document}